**Direct observation of water mediated single proton transport between hBN surface defects**

Jean Comtet[1,*], Benoit Grosjean[2], Evgenii Glushkov[1], Ahmet Avsar[3,4], Kenji Watanabe[5], Takashi Taniguchi[5], Rodolphe Vuilleumier[2], Marie-Laure Bocquet[2] and Aleksandra Radenovic[1,*]

[1]Laboratory of Nanoscale Biology, Institute of Bioengineering, School of Engineering, École Polytechnique Fédérale de Lausanne (EPFL), 1015 Lausanne, Switzerland
[2]PASTEUR, Département de chimie, École normale supérieure, PSL University, Sorbonne Université, CNRS, 24 Rue Lhomond, 75005 Paris, France.
[3]Electrical Engineering Institute, École Polytechnique Fédérale de Lausanne (EPFL), Lausanne, Switzerland.
[4]Institute of Materials Science and Engineering, École Polytechnique Fédérale de Lausanne (EPFL), Lausanne, Switzerland
[5]National Institute for Materials Science, 1-1 Namiki, Tsukuba 306-0044, Japan
[*]jean.comtet@gmail.com, aleksandra.radenovic@epfl.ch

**Aqueous proton transport at interfaces is ubiquitous and crucial for a number of fields, ranging from cellular transport and signaling[1–4], to catalysis[5–8] and membrane science[9–15]. However, due to their light mass, small size and high chemical reactivity, uncovering single proton surface transport at room temperature and in aqueous environment has so far remained out-of-reach of conventional atomic-scale surface science techniques, such as STM[5,16–19]. Here, we use single-molecule localization microscopy techniques to resolve optically the transport of individual excess protons at the interface of hexagonal boron nitride crystals and aqueous solutions at room temperature. Our label-free approach relies on the successive protonation and activation of optically active defects at the surface of the crystal allowing us to resolve interfacial proton transport at the single molecule scale with nanometric resolution and over micrometer range. Proton trajectories are revealed as a succession of jumps between proton-binding defects, mediated by interfacial water. We demonstrate unexpected interfacial proton mobility under illumination, limited by proton desorption from individual defects. The proposed mechanism is supported by *ab initio* molecular dynamics simulations of defected and pristine hBN/water interface. Our observations provide direct experimental evidence at the single molecule scale that interfacial water provides a preferential pathway for lateral proton transport. Our findings have fundamental and general implications for water-mediated molecular charge transport at interfaces.**

Proton transport in bulk water is known to occur via the so-called Grotthuss mechanisms[20], whereby protons tunnel between individual water molecules along liquid wires formed by hydrogen bonds. This remarkable transport mechanism, speculated almost 200 years ago explains the anomalous and peculiarly high mobility of hydronium and hydroxide ions in bulk water[21]. At interfaces, the situation is much more complex, with experimental and theoretical efforts pointing to a wealth of effects, ranging from specific proton desorption barrier[22] potentially facilitated by interactions with water molecules[5] and hydrogen bonding[19,23,24], peculiar charging effects due to water negative self-ion[25], to 2D confinement of protons at hydrophobic interfaces, leading to facilitated lateral transport[1–4]. However, interfacial transport of protons, and its relation with the surrounding aqueous water environment has so far remained elusive, due to the lack of direct measurements at the single molecule scale and in environmental conditions. A finer molecular understanding of proton transport at interfaces would have fundamental importance for a range of fields and materials, from cell membranes in biology[1–4], metallic and oxide surfaces for catalysis and surface science[5–8], to polymeric surfaces for fuel cells[9–11,26,27] and membrane science[12–15].

Here, we use single molecule localization microscopy to resolve the transport of individual excess protons between defects at the surface of hexagonal boron nitride (hBN) crystals. Our label-free approach relies on the protonation-induced optical activation of defects at the surface of the flake. Building upon the recent application of super-resolution microscopy to hBN defects[28,29], we are able to follow spatial trajectories of individual excess protons, through successive hopping and activation of surface defects. We reveal heterogeneous water mediated proton mobility under illumination, with proton transport limited by desorption from individual defects. Our observations demonstrate that interfacial water provides a preferential pathway for proton and charge transport. This finding, along with the chemical nature of the defects in aqueous conditions, are corroborated by full quantum molecular dynamic simulations of pristine and defected hBN/water interfaces. Our findings and observations have general implications for proton transport between titrable surface groups or surface traps, as arising at a variety of biological[1,3,30] and solid-state[6,9,11,17,22] interfaces.

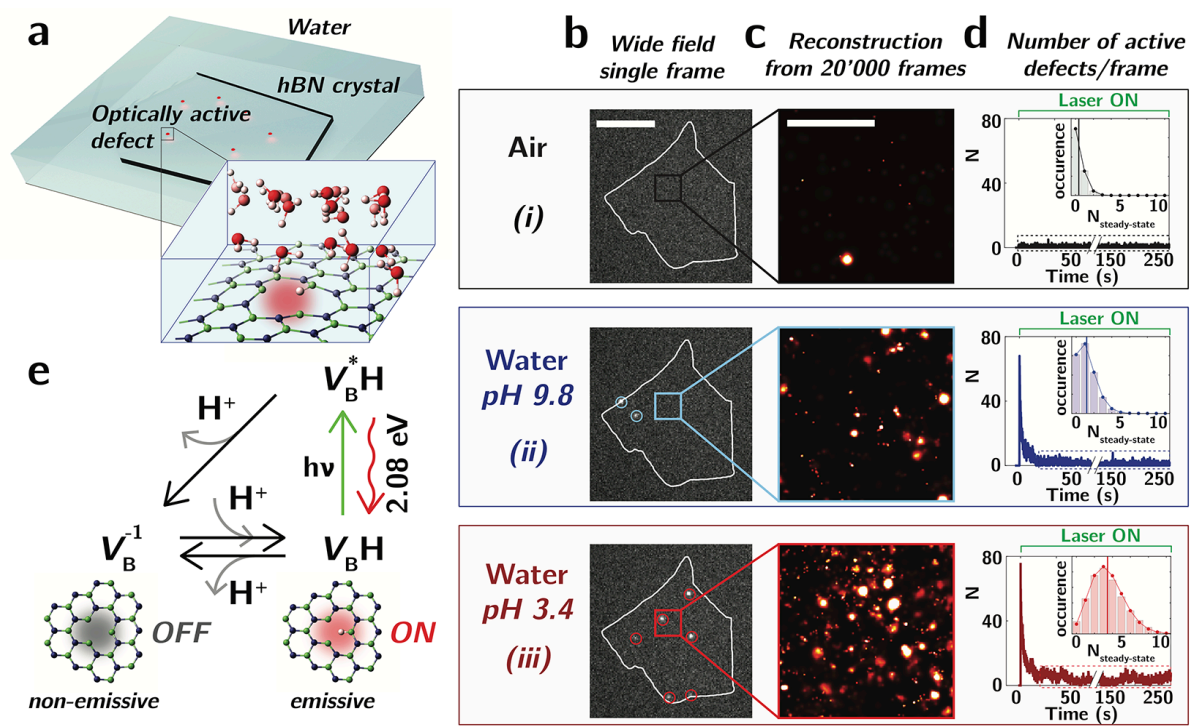

**Figure 1. Reactivity of hBN defects in aqueous conditions: protonation activates defects. (a)** hBN crystal containing irradiation-induced surface defects is illuminated by a continuous green laser, leading to localized emission from optically active defects (red). The crystal can be exposed to various environmental conditions (air or water solutions of varying acidity). Inset: Zoom on a surface defect (protonated boron vacancy $V_B H$), surrounded by water molecules. Boron, nitrogen, oxygen and hydrogen atoms are represented respectively as blue, green, red and white. **(b-d)** Larger defect reactivity in aqueous acidic conditions, comparing flake in air *(i)*, in basic *(ii, pH 9.8)* and acidic *(iii, pH 3.4)* water solutions. **(b)** Wide field image of an hBN flake, obtained with 20 ms exposure time. Emission from individual surface defects can be localized with ~ 5-40 nm uncertainty (red and blue circles, see *SI S1.2*). Scale bar: 5 μm. **(c)** Reconstructed super-resolved images of the flake surface (*SI S1.3*). Scale bar: 1 μm. **(d)** Number $N$ of emissive defects as a function of time for the three environmental conditions. In air, the number of active defects is consistently small. In water, we observe a large number of active defects upon illumination, which decreases to reach a steady-state (dashed box, see *SI S2* and *Fig. S5* for analysis of the relaxation kinetics). Inset shows the histogram of the number of active defects per frame at steady-state $N_{\text{steady-state}}$, fitted by Poisson distributions. Vertical lines in inset show the average number of defects per frame. A larger activity of surface defects is evidenced in acidic conditions, consistent with protonation induced activation of surface defects. **(e)** Three state model for the protonation induced transition between non-emissive negatively charged boron vacancy $V_B^-$, and emissive neutral protonated boron vacancy $V_B H$, with excited state $V_B^* H$ (Fig. S4).

As shown in Figure 1a, our sample is composed of multilayer boron nitride flakes, exfoliated from high quality crystals[31]. Such exfoliated hBN flakes are atomically smooth and host very few intrinsic defect sites[32]. Defects are deterministically induced at the surface of the flake through a brief low-power oxygen plasma treatment[29,33] (*SI S1.4*). Wide-field illumination of the sample with a continuous green laser ($\lambda_{\text{exc}} = 565$ nm) leads to localized emission from optically active defects at the surface of the flake, characterized by uniform emission at 585 nm (2.08 eV), consistent with previous reports[29,34] (*Fig. S1*).

Remarkably, as shown in Figs. 1b-d, we observe a drastic change in the photoluminescence response of the hBN flake when exposed to air (*i*, black) or to aqueous solutions of varying acidity (*ii* blue; *iii* red, corresponding to pH 9.8 and pH 3.4), pointing to the high reactivity of surface defects in aqueous conditions. Fig. 1b shows the wide field image of the hBN flake under uniform illumination, with the flake physical boundary highlighted as white contour. For each frame, emission originating from surface defects can be localized with subwavelength uncertainty, with a typical localization uncertainty $\sigma_{\text{LOC}} \approx 5 - 40$ nm, scaling as $\sigma_{\text{LOC}} \sim \sigma_{\text{PSF}}/\sqrt{N_\phi}$, with $\sigma_{\text{PSF}} \approx 150$ nm fixed by the point spread function of the microscope and $N_\phi$ the number of emitted photons (Fig. 1b, black, blue and red circles, *SI S1.2*). Consecutive localization of emitters over successive frames (here 20'000) allows us to reconstruct a super-resolved spatial map of the defects at the surface of the crystal[28,29], with a zoom on a 2x2 µm$^2$ area shown in Fig. 1c. We observe in Fig. 1c that while only few defects are active in air, a large number of defects are homogeneously activated in aqueous conditions (*Fig. S9* for super-resolved maps of the entire flake). This difference is also highlighted by monitoring the number *N* of emissive defects per frame under illumination (Fig. 1d, inset). In air, the number of active defects per frame is consistently low, with *<N>* $\approx 0.3$ active defects/20 ms. Immersing the flake in water, we observe upon illumination an initially very large number of active defects (here $N \approx 70$, corresponding to $\approx$ 1 active site/µm$^2$), pointing to the activation of defects luminescence due to solvent molecules (*Fig. S9*). The number of active defects decreases upon illumination over tens of seconds (Fig. S5), to reach a steady-state, characterized by *<N>* $\approx 1 - 4$ active defects/20 ms. Importantly, the luminescent state of defects is recovered over sufficiently long dark periods as well as through successive drying and wetting steps (*Figs. S6, S7*).

Varying water acidity further allows us to identify environmental protons $H^+$ as being the chemical species responsible for the activation of defect luminescence. Comparing the two pH conditions in Figs. 1b-d, we indeed observe an increase by a factor of $\approx 2$ in the number of emissive defects at acidic pH (Fig. 1d, inset), as well as an increase in the density of activated emitters (Fig. 1c). This monotonic increase in defect's activity in acidic conditions was systematically observed in all investigated crystals over 12 pH units (*Fig. S10*).

To rationalize our observations, we perform ab-initio molecular dynamic simulations of a defective hBN interface in water (Fig. 1a, inset, *SI S5* and *Supplementary Movie 1*). Based on recent simulations on anhydrous bulk hBN defects[35] and high-resolution TEM images of hBN monolayers[28,36], we probed the reactivity of boron monovacancy complexes in water, namely $V_B H$ (identified as the likely 2 eV emitter[35]), and $V_B^-$ (non-emissive at 2 eV, having a lower acceptor defect state in the gap). Our simulations demonstrate that solvated aqueous protons behave like charge-compensating centers and incorporate easily on the negatively charged defect $V_B^-$, through $V_B^- + H^+ \Rightarrow V_B H$, consistent with the large activation of luminescence observed in the aqueous environment. These numerical observations allow us to propose the phenomenological three-state model, depicted in Fig. 1e. In the absence of illumination, the number of protonated defects is fixed by acid-base equilibrium $V_B^- + H^+ \Leftrightarrow V_B H$. Probing this equilibrium experimentally, we determine a pKa $\geq 14$, consistent with the strong basicity evidenced by the simulations (*Fig. S8*). Upon illumination, defects in the protonated state are converted to their excited state $V_B^* H$, from which they can either radiatively decay back to the ground state $V_BH$, or lose their protons to be converted back to $V_B^-$, through an excited state proton transfer[37] (Fig. S4). This second non-radiative pathway leads to the initial decrease of the number of active defects upon illumination (Fig. 1d and Fig. S5) by effectively shifting the chemical equilibrium between $V_B^-$ and $V_BH$, reaching a second steady-state level under constant laser illumination. This photoacidic behavior is consistent with the relative excited state energy levels of the protonated and deprotonated defect[35] (Fig. S4). Consistently, a decrease in the number of active defects at steady-state is observed for increasing illumination power (*Fig. S11*). While reported here for the first time for defects in hBN, this ON/OFF transition between distinct protonation states is commonly observed for fluorescent dyes[38,39]. Note finally, that while our observations and ab-initio simulations are consistent with the $V_BH/V_B^-$ transition between emissive and non-emissive states, other types of defects with nitrogen dangling bonds and distinct protonation states could also be responsible for these observations.

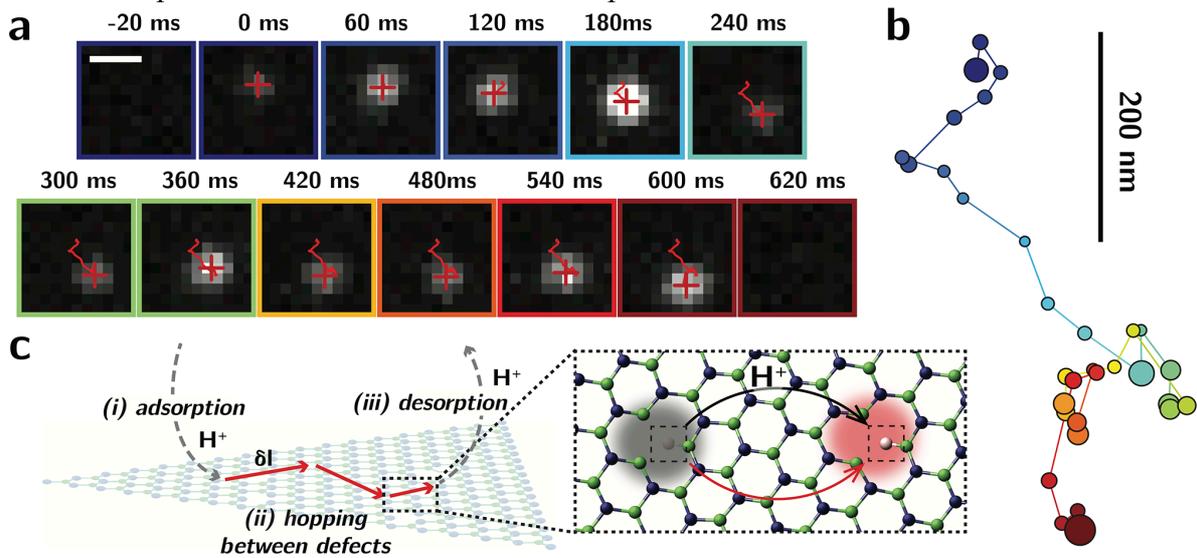

**Figure 2. Luminescence migration reveals proton trajectories.** (a) Time series for spatial migration of luminescence at the surface of the flake, with wide field images showing localized diffraction-limited spot at the surface of the flake (red cross) and reconstructed spatial trajectory in red (*Supplementary Movie 3*). Scale bar: 500 nm. Projected pixel size is 100 nm. (b) Reconstructed trajectories showing successive activation of adjacent defects at the surface of the flake and color-coded with increasing time. Localized defects are represented as dots, with radius corresponding to localization uncertainty. (c) Schematic depicting luminescence migration events, consisting of successive *(i)* proton adsorption (appearance of a luminescence spot at the surface of the flake) *(ii)* excess proton hopping between surface defects (diffusion of the luminescence spot) and *(iii)* proton desorption from the surface of the flake (disapearance of the luminescence spot).

Since the defects are emissive in their protonated form, monitoring luminescence events at the flake's surface allows us to directly track down the spatiotemporal dynamics of defects protonation. Fig. 2a shows the temporal evolution over 600 ms of the luminescence in a 1x1 μm² region at the surface of a flake immersed in DI water (pH ~ 5.5), with subpixel localization of the emitter's position shown as the red cross. As highlighted in this sequence of images, a single diffraction-limited luminescence spot spatially wanders between successive frames over a total distance of ≈ 500 nm. Consecutive localizations allow us to reconstruct the position of successively activated defects, shown as the red line trace in Fig. 2a and the reconstructed trajectory in Fig. 2b, with radius corresponding to uncertainty in localizations. The observation of the consecutive activation of luminescence of nearby defects over 30 successive frames, points to the presence of a single activating excess proton, hopping from defects to defects (Fig. 2c, inset) and leading to the observed spatiotemporal activation of luminescence. Importantly, monitoring defect activation over the whole flake allows us to discard artefacts in these observed trajectories related to stage drift or random activation of emitters (*Figs. S14* and *S15*). As schematically represented in Fig. 2c, this observed sequence of correlated luminescence events must then corresponds to *(i)* the adsorption of a proton at one defect's site, leading to the appearance of a luminescence spot at the flake's surface (Fig. 2a, *t* = 0 ms) followed by *(ii)* the hopping of the excess proton between nearby surface defects over the total residence time $T_R$, with successive hopping length $\delta l$ and *(iii)* the desorption of the proton from the flake surface, leading to the extinction of luminescence (Fig. 2a, *t* = 620 ms). Importantly, we demonstrate through simulations that such correlated luminescence events and trajectories cannot stem from the random activation of emitters at the surface and must correspond to the correlated transfer of a single excess proton between defects (*SI S3.4*). Following the three-state model of Fig. 1e, the variation in luminescence intensity observed between successive frames could stem from fluctuations between radiative and non-radiative recombination of the excited defects, e.g. due to transient proton unbinding and geminate recombination[37]. Albeit in a different context, the concepts presented here for single proton tracking are similar to strategies explored in single molecule and single enzyme catalysis[40–42].

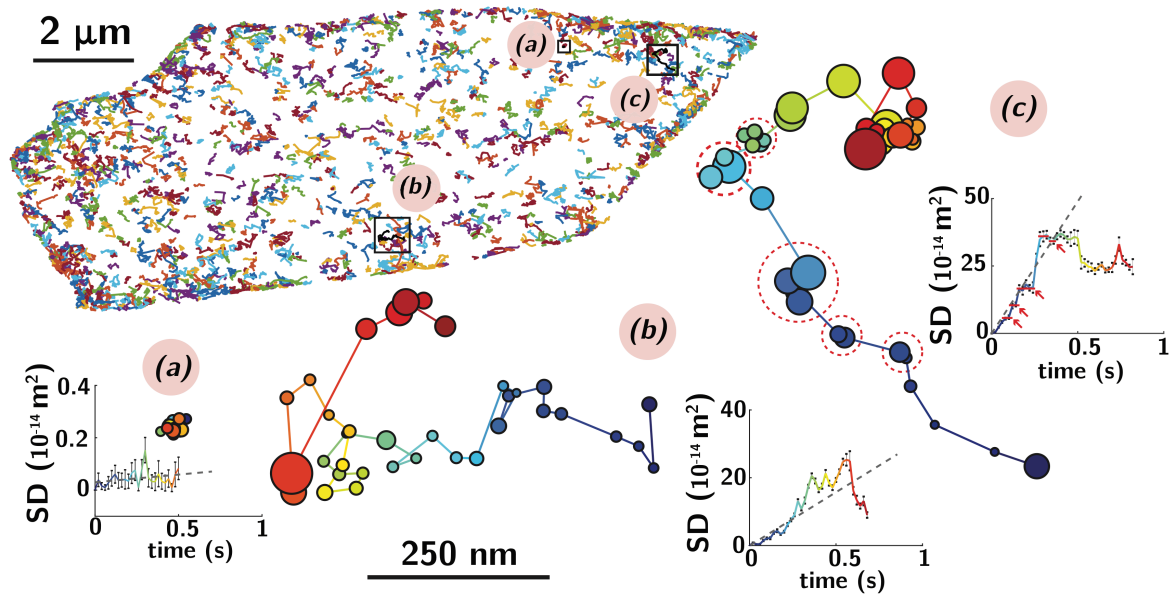

**Figure 3: Large-scale mapping of proton trajectories.** Trajectories longer than 10 frames (200 ms) measured at the surface of the flake. *(a-c)* Highlighted representative trajectories, with the corresponding evolution of the mean square displacement MSD (*Supplementary Movies 4-6*). Defect positions are rendered as circles with radius corresponding to the localization uncertainty. SD is square displacement. Black dashed line is a linear fit of the SD for the first 300 ms. In *(c)*, dashed red circles in trajectory and red arrows in SD show adsorbing steps at some defect sites.

These correlated proton trajectories occur consistently and repeatedly upon constant illumination (*Supplementary Movie 2*). We track and analyze their dynamics using standard single particle tracking techniques[43], focusing on the steady-state regime with a constant averaged density of active defects per frame of ≈0.1-1 per 10 μm² (Fig. 1d, dashed boxes). Individual trajectories are defined by correlating localizations less than ~300 nm apart between successive time frames. This threshold is rigorously defined by measuring the statistic of hopping length $\delta l$ (*Fig. S13*). Importantly, this tracking methodology is robust with respect to the correlation length and sampling time and is validated against simulations of random activation of emitters (*Fig. S14*). As shown in Fig. 3, ≈ 1700 individual trajectories longer than 200 ms (10 successive frames) can be successfully identified over 180 s (*Supplementary Movie 2*). Representative trajectories are highlighted in Fig. 3*a-c* (see also *Fig. S19* and *Supplementary Movies 4-8*). Remarkably, a large heterogeneity is observed between distinct trajectories at the single molecule scale. Some excess protons remain at a fixed position *(a)*, while others migrate over up to 1 μm (Fig. 3, *b – c* and *Fig. S19*). Long adsorbing steps (40-100 ms) within one uncertainty-limited defect, separated by relatively long hopping events (50-200 nm) are also observed in some trajectories (red dashed circles, *c*). For each individual trajectory, we can compute the associated Square Displacement $SD(t) = (X(t) - X(t=0))^2$ which characterizes the diffusive character of these random walks. From the initial increase of the square displacement with time, one can extract a diffusion coefficient $D$ for each individual trajectory, as $SD = 4.D.t$ (dashed line in the SD graphs), which is found to be respectively $D_{(a)} \approx 10^{-16}$ m²s⁻¹ (no diffusion), $D_{(b)} \approx 25 \cdot 10^{-14}$ m²s⁻¹ and $D_{(c)} \approx 8 \cdot 10^{-14}$ m²s⁻¹. Note the larger number of observed trajectories at the edges migth be due to a larger density of defects (see *Fig. S27*).

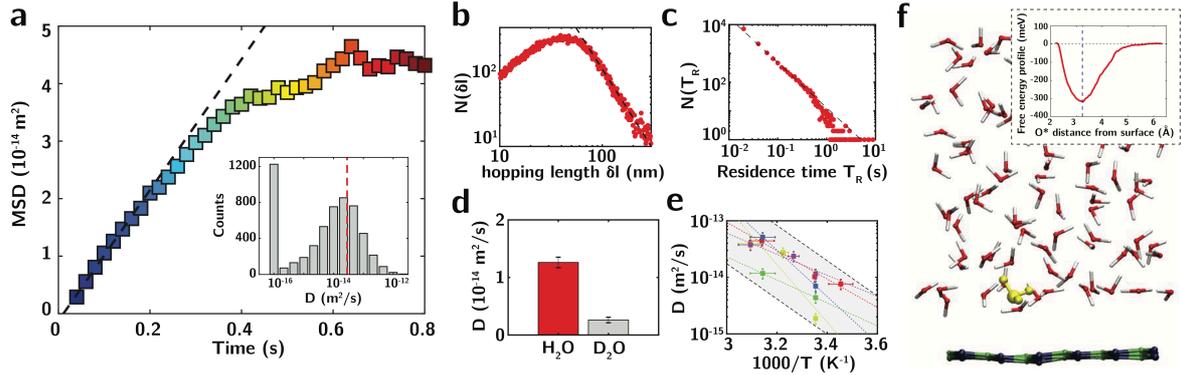

**Figure 4: Mobility and segregation of protons at interfaces. (a)** Variation of the averaged mean square displacement MSD as a function of time for the flake in Fig. 3. Dashed line is a linear fit from which we extract the averaged diffusion coefficient $D = 2.8 \times 10^{-14}$ m²s⁻¹. Inset shows the distribution of the diffusion coefficient determined on individual trajectories, with vertical dashed line the averaged diffusion coefficient. **(b)** Distribution of step length $\delta l$ between successive jumps. Dashed line is power-law fit with exponent $\nu \approx 2.6$. **(c)** Distribution of residence time $T_R$. Dashed line is power-law fit with exponent $\mu \approx 1.6$. **(d)** Isotope effect, comparing diffusion coefficient in D₂O and H₂O in a distinct flake. Error bars represent standard deviation in diffusion coefficient. **(e)** Variation of diffusion coefficient as a function of inverse temperature for 5 distinct flakes, with linear fits shown as dashed colored lines. Black dashed lines are visual guides showing activated Arrhenius behavior with the mean activation energy of 0.62 eV (see Fig. S28). **(f)** Simulation snapshot of the trajectory of an aqueous hydronium ion physisorbed at the pristine hBN/water interface (See *Supplementary Movie 9*). Inset shows the computed free energy profile of the hydronium ion as it approaches the hBN layer, with a physisorption well of -0.3 eV centered around the maximum of water density at 3.3 Å (blue vertical dotted line in inset, see *SI S5*).

As shown in Fig. 4a, we characterize the interfacial mobility of protons at the surface of the flake through the evolution of the Mean Square Displacement MSD averaged over all observed trajectories (the averaged $MSD = <(X(t) - X(t=0))^2>$ is a well-defined quantity, independent of sampling time and tracking parameters, see *SI S3*). As shown in Fig. 4a, the MSD follows an initial linear increase, characteristic of a standard diffusive behavior at short times ($t < 300$ ms), followed by a sub-diffusive behavior at longer times, possibly due to longuer adsorbing events at some defect's sites. The linear regime allows us to define the average diffusion coefficient $D = 2.8 \times 10^{-14}$ m²s⁻¹ and typically found of the order of $10^{-14}$ m²s⁻¹ for the majority of flakes. We further show in the inset the broad distribution of diffusion coefficients from individual single-molecule trajectories, with a significant proportion of trajectories characterized by no net observed motion ($D < 1.10^{-16}$ m²s⁻¹, as in Fig. 3a). In order to analyze in more details the statistics of these bi-dimensional proton walks, we plot respectively in Figs. 4b and 4c the distribution of hopping length $\delta l$ and residence time $T_R$ at the surface (calculated as $T_R = N_R \Delta t$, with $N_R$ the trajectory length in frames and $\Delta t = 20$ ms the sampling time). Those distributions are well approximated by power laws $N(\delta l) \sim \delta l^{-\nu}$ and $N(T_R) \sim T_R^{-\mu}$, with here $\nu \approx 2.6$ and $\mu \approx 1.6$, and which are found typically in the range $\nu \in [2.4 - 4]$ and $\mu \in [1.6 - 2.5]$ (Fig. S13). The power law scaling for the jump length $\delta l$ is reminiscent of Levy-type processes[44,45], and demonstrates the anomalous non-Brownian character of these hopping events, due to the finite distance between randomly distributed defects (defect density on this flake can be estimated to be at least 500 μm$^{-2}$, leading to an averaged interdefect distance of 40 nm). The power-law scaling of the residence time naturally arises from the length of a diffusion-controlled escape process, and is larger than for normal diffusion, for which $\mu = 1.5$. Importantly, a large fraction (~70%) of protons remains on the flakes' surface between each frame, leading to trajectories which are subsequently analyzed (Fig. S20).

The orders of magnitude difference between the measured diffusion coefficient $D \approx 10^{-14}$ m$^2$/s for proton surface transport and the hydronium diffusion coefficient in the bulk[21,37,46] $D_{\text{bulk}} \approx 10^{-8} - 10^{-7}$ m$^2$/s and at biological membranes[2-4] $D_{\text{membrane}} \approx 10^{-11} - 10^{-9}$ m$^2$/s suggests the presence of a strong rate-limiting step for interfacial proton transport. We thus compare the surface transport of the two isotopes - hydrogen and deuterium (Fig. 4d, comparing transport in H$_2$O and D$_2$O in a distinct flake, see also *Fig. S22*). As shown in Fig. 4d, diffusion is hindered in D$_2$O by at least a factor of 4 compared to H$_2$O. This isotopic hindrance to diffusion is larger than the factor 1.5 to 2 that one would expect from hindrance of either Grotthuss-like proton transfer or self-diffusion[47,48], pointing to desorption from defects rather than transport between nearby defects as the rate-limiting step for excess proton transport. Such desorption-limited transport is consistent with the low value of the interfacial diffusion coefficient, the large distribution in diffusion coefficients observed in individual trajectories (Fig. 4a, inset), as well as the long adsorbing steps evidenced at some defects' sites (Fig. 3, trajectories *a* and *c*). Note finally that this hindered desorption-limited transport validates *a posteriori* the choice of our spatiotemporal resolution ($\Delta t = 10 - 20$ ms and $\Delta l = 300$ nm) for which the largest measurable diffusion coefficient is $\Delta l^2/4\Delta t \sim 10^{-12}$ m$^2$/s. This value, despite being smaller than the bulk hydronium diffusion coefficients, remains two orders of magnitude larger than the desorption-limited interfacial diffusivities, allowing us to consistently measure and caracterize the hindered interfacial proton transport between defects (*SI S3*). Additional characterization of the proton transit time between defects (corresponding to unhindered interfacial proton transport) would require at this 100 nm scale a temporal resolution $\delta t \approx 1$ μs, unreachable with current state-of-the-art single molecule tracking techniques[49].

Based on these insights, we can, for desorption-limited transport, express the diffusion coefficient as $D = \frac{1}{4}\Gamma a^2$, with $\Gamma$ [s$^{-1}$] a jump rate and $a$ the characteristic jump length between nearby defects. This jump rate scales as $\Gamma \sim \nu \exp\left(\frac{-\Delta F}{kT}\right)$, with $\nu$ [$s^{-1}$] a molecular frequency and $\Delta F$ a free energy desorption barrier from the defect[50]. As an order of magnitude estimate, we take the attempt frequency as $\nu \sim 1/\tau$ with $\tau$ of the order of nanoseconds corresponding to the excited state lifetime[51] (Fig. S2) and $a \approx 10 - 100$ nm, leading to a typical desorption energy barrier of 16-20 k$_B$T $\approx$ 0.4-0.5 eV. As shown in Fig. 4e, we indeed observe an increase in proton mobility with increasing temperature, from which we extract a mean activation energy $\Delta E \approx 0.62 \pm 0.12$ eV, demonstrating the predominantly enthalpic nature of this free-energy barrier (Fig. S28). This barrier then characterizes the energy necessary to break the NH covalent bond from the excited defect $V_B H^*$ and for the solvated $H^+$ to escape the electrostatic attraction of the negatively charged vacancy $V_B^-$. Consistently, this barrier is much smaller than the hydrogen removal energy > 2.34 eV predicted to break the NH bond from the $V_B H$ defect in gas phase[35], as the proton desorption barrier might be reduced here by the presence of nearby hydrogen accepting water molecules, as well as by the laser irradiation. Indeed, we did not observe any proton mobility in air (*Fig. S26*), despite the presence of adsorbed water at the flake surface in ambient conditions (40% relative humidity), demonstrating the crucial role of bulk water in mediating proton mobility at the flake's surface and consistent with recent simulations[23]. Desorption-limited transport is furthermore confirmed by the weak dependence of mobility on illumination power (*Fig. S21*). Furthermore, as shown in *Figs. S23-S24*, the presence of salt and dissolved gas does not affect significantly the interfacial proton mobility, while we observed in a majority of flakes a net increase in mobility at low pH (*Fig. S25*), consistent with increased defect activity and change of surface state. Finally, we note that the high purity and atomic flatness of the hBN surfaces should also lead to reduced probability of proton trapping at non-emissive sites, allowing the direct observation of excess proton transport between nearby defects.

The emerging picture is thus that of a desorption-limited transport of protons between surface defects, mediated by interfacial water. While several experiments have reported evidence for interfacial proton mobility at surfaces through ensemble measurements[1–4,52], the trajectories observed here at the flake's surface (Figs. 2-3) represent the first direct observation at the single-molecule scale of the interfacial segregation of proton excess at the solid-water interface, demonstrating that interfacial water provides a preferential pathway for charge transport. Indeed, in the absence of any free-energy barrier trapping protons at the interface, a proton desorbing into the bulk would irreversibly diffuse over $\delta l \sim \sqrt{D_{\text{bulk}} \cdot \delta t} \approx 300$ μm during the $\delta t = 20$ ms sampling time, preventing any correlations in the activation of defects $\approx 100$ nm appart. The power-law tail of the surface residence time (Fig. 4C), concomitant with the finite diffusivity (Fig. 4A) accordingly demonstrates the large probability of near-surface charges to remain segregated and mobile at the surface. In-plane proton transport at the flake surface (Fig. 2b, step *ii*) must thus be strongly favored compared to proton desorption into the bulk (Fig. 2b, step *iii*), due to the presence of an interfacial free-energy barrier leading to the segregation of the excess protons at the hBN/water interface.

To probe the segregation of interfacial protons in more details, we simulated the dynamics of a hydronium ion at the interface of water and pristine hBN (Fig. 4f). As observed in this 20 ps trajectory (*Supplementary Movie 9*), the hydronium ion indeed remains segregated (physisorbed) at the interface, while keeping high lateral mobility through grotthus transfer, with a lateral diffusion coefficient $D \approx 8.10^{-9}$ m$^2$.s$^{-1}$, close to the bulk hydronium diffusion coefficient[21,37,46] $D_{\text{bulk}} \approx 10^{-8} - 10^{-7}$ m$^2$.s$^{-1}$. As shown in the inset of Fig. 4f, the computation of the free-energy of the aqueous hydronium approaching a pristine hBN surface further confirms the presence of an interfacial -0.3 eV physisorption well (*SI S5*). Mechanistically, several effects can be invoked to explain the observed affinity of protons to interfaces. First, in hydronium ions, assymetric charge distribution leads to an amphiphilic surfactant-like character[4,53], which could lead to segregation due to the hydrophobic nature of the hBN interface. Second, the ionic nature of the insulating hBN crystal[54] could also be responsible for electrostatic trapping of the positively charged hydronium ion. Third, as hydronium donates three hydrogen bonds to water it leads to straining and disruption of the hydrogen bonding network[55]. This effect is reduced at interfaces - at which the hydronium oxygen tend to point away from water[48]- and could lead to trapping of the ion[2,4,52,55].

The combination of super-resolution microscopy and single particle tracking on hBN surface defects allowed us to reveal proton trajectories between adjacent defects at the single-molecule scale. These observations establish that interfacial water provides a preferential pathway for proton transport, with excess proton diffusing along the surface rather than through the bulk, leading to the observed spatiotemporal correlations in the activation of nearby defects. The direct observation of this interfacial proton pathway has broad implications for charge transport in a range of fields and materials, and suggests to tune defects' densities, binding affinities and illumination to optimize and control interfacial proton transport. Our experiments thus represent a promising platform for the investigation of proton transport at the single molecule scale, opening up a number of avenues, e.g. related to the interplay of flow or confinement with molecular charge transport at liquid/solid interfaces.

**Data Availability**
The data that support the findings of this study are available from the corresponding authors on reasonable request.

**Acknowlegment**
JC acknowledges valuable discussions with Adrien Descloux and Vytautas Navikas.

**Funding Sources**
This work was financially supported by the Swiss National Science Foundation (SNSF) Consolidator grant (BIONIC BSCGI0_157802) and CCMX project ("Large Area Growth of 2D Materials for device integration"). E.G. acknowledges support from the Swiss National Science Foundation through the National Centre of Competence in Research Bio-Inspired Materials. The quantum simulation work was performed on the French national supercomputer Occigen under the DARI grants A0030807364 and A0030802309. MLB acknowledges funding from ANR project Neptune. K.W. and T.T. acknowledge support from the Elemental Strategy Initiative conducted by the MEXT, Japan and the CREST(JPMJCR15F3), JST.